# Evidence-Based Filters for Signal Detection: Application to Evoked Brain Responses

M. Asim Mubeen[a], Kevin H. Knuth[a,b,c]

[a]Knuth Cyberphysics Lab, Department of Physics, University at Albany, Albany NY, USA
[b]Department of Informatics, University at Albany, Albany NY, USA
[c]Autonomous Exploration Inc., Andover MA , USA

**Abstract.** Template-based signal detection most often relies on computing a correlation, or a dot product, between an incoming data stream and a signal template. Such a correlation results in an ongoing estimate of the magnitude of the signal in the data stream. However, it does not directly indicate the presence or absence of the signal. The problem is really one of model-testing, and the relevant quantity is the Bayesian evidence (marginal likelihood) of the signal model. Given a signal template and an ongoing data stream, we have developed an evidence-based filter that computes the Bayesian evidence that a signal is present in the data. We demonstrate this algorithm by applying it to brain-machine interface (BMI) data obtained by recording human brain electrical activity, or electroencephalography (EEG). A very popular and effective paradigm in EEG-based BMI is based on the detection of the P300 evoked brain response which is generated in response to particular sensory stimuli. The goal is to detect the presence of a P300 signal in ongoing EEG activity as accurately and as fast as possible. Our algorithm uses a subject-specific P300 template to compute the Bayesian evidence that a applying window of EEG data contains the signal. The efficacy of this algorithm is demonstrated by comparing receiver operating characteristic (ROC) curves of the evidence-based filter to the usual correlation method. Our results show a significant improvement in single-trial P300 detection. The evidence-based filter promises to improve the accuracy and speed of the detection of evoked brain responses in BMI applications as well the detection of template signals in more general signal processing applications

## INTRODUCTION

Neurophysiology in general, and brain machine interface (BMI) in particular, relies on the detection and characterization of the electric and magnetic field potentials produced by the brain in response to sensory stimulation or in association with its cognitive and/or motor operations and planning. These potentials originate from the transmembrane current flow produced by multiple ensembles of hundreds of thousands of synchronously firing neurons. Human scalp electroencephalographic (EEG) recording has the advantage of being noninvasive, inexpensive and portable, which make it a very popular technique among the BMI community.

Event Related Potentials (ERPs) are evoked brain responses synchronized to sensory, cognitive and motor events. As such, they consist of relatively reproducible waveshapes embedded in background EEG activity [10, 12]. In particular the P300 evoked potential [9, 17] is a positive peak that is evoked 300 ms stimulus onset. The P300 is widely used in Brain Computer Interface (BCI) applications, but recent studies shows it is also useful in the diagnosis of neurological disorders and lie detection. Despite its wide use, detection of the P300 component in single trial recordings is still a challenge. More often, multiple trials are needed to detect the P300 component in ongoing EEG activity, which decreases the overall speed of the BMI system.

In general, the detection of an ERP is made difficult by its low signal-to-noise (SNR) ratio compared to the ongoing background EEG. The most commonly-used method to estimate the ERP stimulus is coherent averaging (CA), which averages a large number of time-locked epochs of the identical stimuli presented to the subject. CA is used in online applications of BCI [3, 5] however this technique suffers from several drawbacks [13,16]. One drawback is that CA implicitly assumes that the ERP waveshape is identical from trial-to-trial. This is known to be a poor assumption and, in fact, this feature is utilized in some advanced source separation methods [4, 10, 18].

Several detection methods work by correlating a template signal with the ongoing EEG. For example, the Woody filter [19] takes the dot product of the ongoing EEG signal with the signal template. Other classification techniques like Pearson's correlation method (PCM), Fisher's linear discriminant (FLD), stepwise linear discriminant analysis (SWLDA), linear support vector machine (LSVM), and the Gaussian kernel support vector machine (GSVM) have been presented and compared [11, 15]. These methods that convert the problem of detecting P300 into a binary search problem (P300 present or not-present) have been widely used in BCI applications. In this paper we present a Bayesian evidence-based technique to detect the P300 response in ongoing EEG. This technique is not limited to EEG signals and can be used for any kind of signal to be detected. We demonstrate the technique by applying it to synthetic EEG data to detect brain evoked responses similar to the P300. We compare the results with the template correlation (dot-product/Woody filter) method and demonstrate the efficacy of our proposed method by comparing the receiver operating characteristic (ROC) curves [2, 14, 20] of each technique.

## EVIDENCE-BASED FILTER

The evidence (marginal likelihood) in the Bayes' theorem can be thought of as quantifying the degree to which the data provides evidence for the model. In our problem, the model consists of the presence of a target template signal (P300) with unknown amplitude. Given a segment of EEG data, the evidence will be high if the data provides evidence for the template signal being present and low otherwise. This effect forms the basis of the evidence-based filter. To deal with an ongoing EEG signal, we compute an ongoing evidence index by working with a sliding window extracted from the ongoing data stream. The computed evidence then can be compared to a threshold value to indicate the presence or absence of the target signal. To utilize this technique, one must first generate a template signal. In the case of P300 response the intra-individual variability [1] requires that we first obtain an individual-specific model of the P300 from EEG data recorded from that subject. The subject-specific P300 can be identified by using available source separation methods, such as differentially Variable Component Analysis (dVCA) [10]. This method takes into account the trial-to-trial variability of amplitude, latency and waveshape and estimates the electrode coupling matrix and ERP amplitude distribution. This information then can be used in evidence-based filter to aid in the detection process.

From Bayes' theorem the evidence can be written as

$$Z = P(data \mid I) = \int_{-\infty}^{\infty} P(data \mid model, I) \, P(model \mid I) \, dmodel \tag{1}$$

where, P(data | model, I) is the likelihood and P(model | I) is the prior probability. Our model consists of a target signal template, s(t), that has a single-trial amplitude, α. We model the ongoing stream of data x(t) recorded by the mth detector as

$$x_m(t) = C_m \, \alpha \, s(t) \tag{2}$$

where $C_m$ represents the coupling between the ERP and the mth electrode. This coupling factor is an element of the coupling matrix also known in the source separation literature as the mixing matrix [10].

We assign a Gaussian likelihood for ongoing EEG data window x(t)

$$P(x(t) \mid C, s(t), \alpha, I) = (2\pi\sigma_0^2)^{-MT/2} \exp\left(-\frac{1}{2\sigma_0^2}\left(\sum_{m=1}^{M}\sum_{t=1}^{T}(x_m(t) - C_m \alpha \, s(t))^2\right)\right) \tag{3}$$

and a Gaussian prior probability for the amplitude of the signal in the data

$$P(\alpha \mid I) = (2\pi\sigma^2)^{-1/2} \exp\left(-\frac{1}{2\sigma^2}(\alpha - \hat{\alpha})^2\right) \tag{4}$$

The evidence can be written as

$$Z = \int_{-\infty}^{\infty} P(x(t) \mid C, s(t), \alpha, I) \, P(\alpha \mid I) \, d\alpha \tag{5}$$

Substituting the assigned likelihood and prior, we can write the product of Gaussians in the integrand as a single Gaussian, which can be analytically integrated to give

$$Z = (2\pi\sigma_0^2)^{-MT/2} \times \exp\left[\left(\frac{1}{2\sigma_0^2\sigma^2}\right)\left(\frac{E^2}{D} - F\right)\right] \tag{6}$$

where D, E and F are given as

$$D = \left[\sigma_0^2 + \sigma^2 \sum_{t=1}^{T}\sum_{1}^{M} C_m^2\, s^2(t)\right], \qquad E = \left[\sigma_0^2 \hat{a} + \sigma^2 \sum_{t=1}^{T}\sum_{1}^{M} C_m\, x_m(t)\, s(t)\right]$$

and

$$F = \left[\sigma_0^2 \hat{a}^2 + \sigma^2 \sum_{t=1}^{T}\sum_{1}^{M} (x_m(t))^2\right]$$

It is computationally convenient to work with the logarithm of the evidence

$$\log Z = -\frac{MT}{2}\log(2\pi\sigma_0^2) + \frac{1}{2}\log\left(\frac{\sigma_0^2}{D}\right) + \left(\frac{1}{2\sigma_0^2 \sigma^2}\right)\left(\left(\frac{E^2}{D}-F\right)\right) \tag{7}$$

which is then used as an index for signal detection.

## METHODOLOGY

To evaluate the accuracy and robustness of the evidence-based filter in the presence of noise we generated synthetic EEG signals. We randomly mixed synthetic P300 signals with a synthetic ongoing EEG signal plus noise. The evidence-based filter was applied by sweeping a window containing the synthetic P300 across the simulated EEG activity and calculating the log evidence value for each window. This log evidence value is then used as an index to detect the P300 signal in the EEG activity. The results were compared to those obtained using the correlation (dot-product) method, which was applied by applying the window containing synthetic P300 across the synthetic EEG data. The result of the dot-product was computed and used as a detection index.

To compare the results of both methods we used Receiver Operator Characteristics (ROC) curves [2, 14, 20]. ROC curves are generated by plotting a graph between sensitivity (True Positive Fractions) versus 1 – specificity (False Positive Fractions). Where sensitivity is the proportion of actual targets which are correctly detected as targets and specificity is the proportion of non-targets which are correctly detected as non-targets. The log evidence threshold value is selected by making a graph of sensitivity and specificity versus criterion value (in our case the value of log evidence and the dot-product). The criterion value at which the sensitivity and specificity curves intersect is used as threshold value for signal detection. This threshold value then assures a balance between true positive detections and false negative detections.

### Synthetic EEG Data

We simulated the EEG recorded for seven detectors. For a single channel, we mixed different sine waves ranging from 3-25 Hz. Then we added pink noise (1/f noise) to all the channels of one data set. To check the robustness of our detection method, we used five different noise levels of 20%, 25%, 30%, 35% and 40% of the signal amplitude to produce five different data sets. These result in SNR levels of 14, 12, 10.5, 9 and 8 dB for the five noise levels.

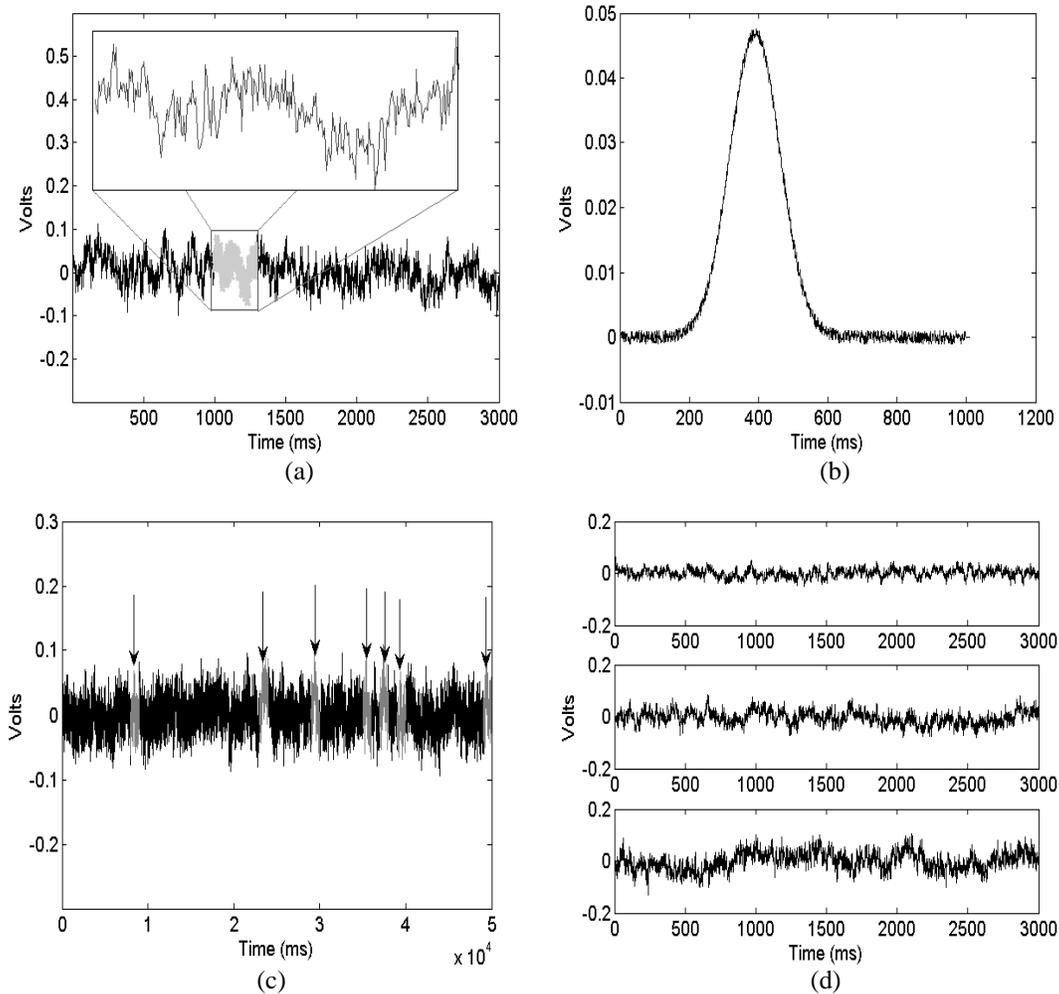

**FIGURE 1.** a) Synthetic EEG data b) Simulated target signal c) Simulated EEG data with target signal in grey and indicated by arrows d) EEG data with noise levels 20%, 30% and 40% from top to bottom.

The synthetic EEG for a single detector is shown in Figure 1a. The synthetic P300 target signal is shown in figure 1b. The target signal is then randomly added to the synthetic ongoing signal at a variety of positions to simulate the responses from a sensory experiment (Figure 1c). The times at which the targets appear in the data are recorded so that we can compute the false negative and false positive detection rates. This is used to plot the ROC curves to compare the performance of our detection method.

## RESULTS

We applied the evidence-based filter to the synthetic EEG data and compared the results with those obtained using the dot-product filter. We computed the Receiver

Operator Characteristics (ROC) curves to compare the results of both methods. ROCs serve as performance measure of signal detection [14, 20] by plotting the true positive rate (also known as *Sensitivity*) versus the false positive rate (1 – *Specificity*). *Sensitivity* can be expressed as the fraction of correct detected targets (P300), and *Specificity* is a fraction of non-targets (Non-P300) detected as non-targets. The area under the ROC curve can be used to compare the detection methods. For the perfect detection method the area under ROC curve will be unity and for a completely failed method it will be zero, so that the more accurate the detection method, the greater the area under the curve. The resulting ROC curves are shown in Figure 2. One can see that the evidence-based method significantly outperforms the dot-product method. At a very high level of noise both methods operate at almost the same efficacy with the evidence-based filter slightly out-performing the dot product.

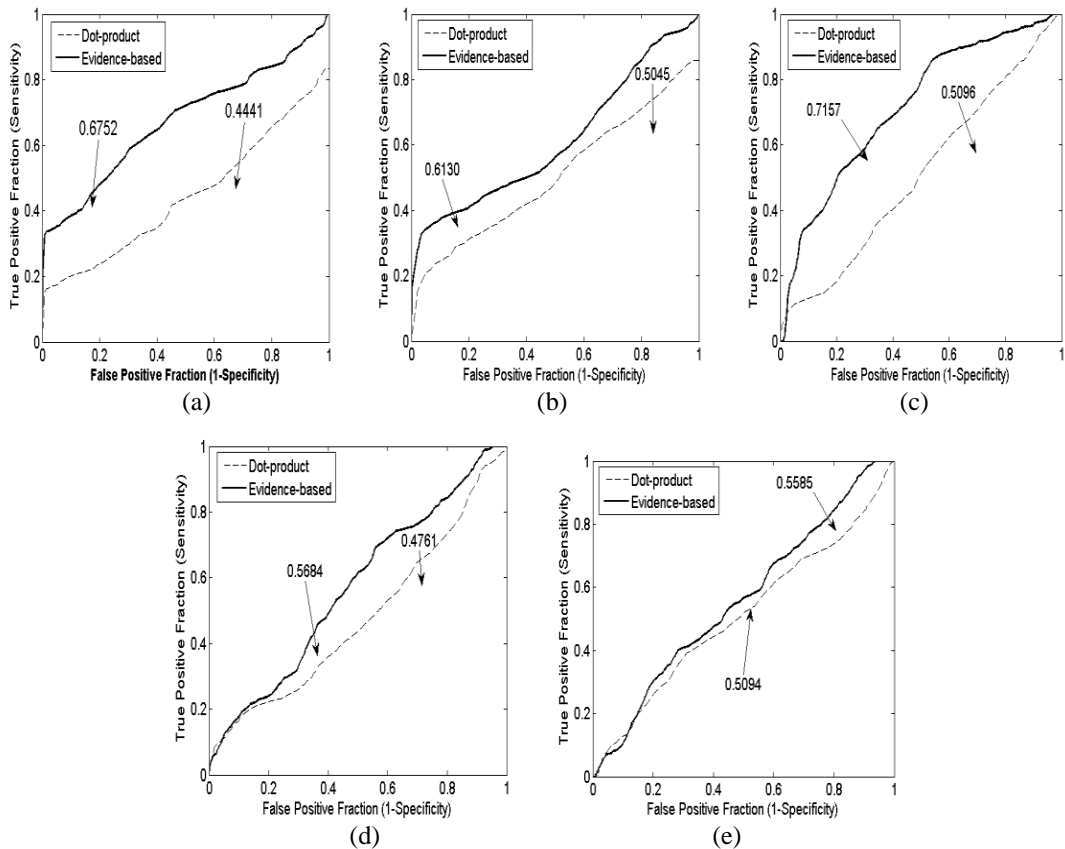

(a) (b) (c)

(d) (e)

**FIGURE 2.** Receiver Operative Characteristic curves for both the dot-product filter (dashed) and evidence-based filter (solid). The area under the curve represents how well the detection method is working. The different panels show the results of noise levels 20%, 25%, 30%, 35% and 40% respectively. In each case the evidence-based filter out-performs the dot-product filter.

## Detection Cutoff

The selection of a detection threshold value is a difficult task. By selecting a higher log evidence threshold value, the false positive fraction will decrease and the specificity will increase, but the true positive fraction and sensitivity will decrease.

With a lower threshold value, the sensitivity will increase but the false positive fraction will also increase. To identify an objective log evidence threshold value (as well as a dot-product threshold value), we graph the criterion value versus the sensitivity and specificity and note that due to the tradeoff, these two curves cross at a single criterion value (Figure 3), which is then used as the detection threshold.

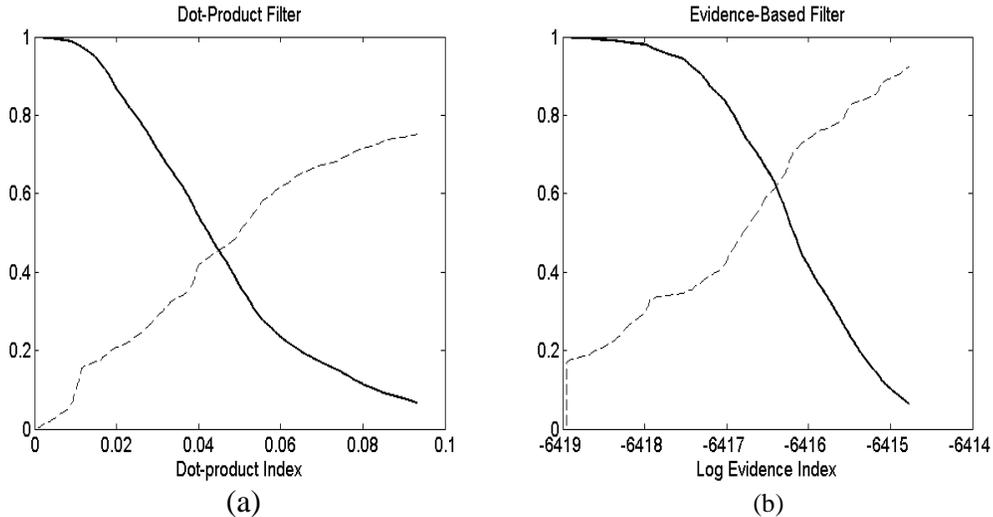

**FIGURE 3.** Graph of criterion value verses Specificity (dashed line) and Sensitivity (solid line) for both methods. The intersection point of two curves can be used as threshold value.

## CONCLUSION

We developed a new detection method where we use Bayesian evidence to detect the presence of a particular signal in the incoming stream of data. We calculate the value of the log evidence by applying a window of the target signal to the incoming ongoing data stream. The calculated log evidence value is then used as an index for identifying the presence of the target signal. To check the robustness of our method we applied the technique to synthetic ongoing EEG mixed with periodically-occurring P300 target stimuli as well as five different noise levels with SNR values 14, 12, 10.5, 9 and 8 dB, respectively. We compare our method with the dot-product (template correlation /Woody filter) method. To compare both methods we used Receiver Operator Characteristics (ROC) curves. The area under ROC curves indicate that the evidence-based filter works better than dot-product (template correlation /Woody filter) method.

## ACKNOWLEDGMENTS

A. Mubeen would like to thank the MaxEnt 2011 organizers and Department of Physics at University at Albany for support provided to attend the conference. The authors would like to thank Dr. Dennis J. McFarland for valuable discussions, access to EEG data and instruction regarding the BCI2000 system.